# Compact gauge $K$ vortices

C. Adam[a*], P. Klimas[a**], J. Sánchez-Guillén[a†], and A. Wereszczyński[b††]

[a)] Departamento de Fisica de Particulas, Universidad de Santiago
and Instituto Galego de Fisica de Altas Enerxias (IGFAE)
E-15782 Santiago de Compostela, Spain
[b)] Institute of Physics, Jagiellonian University,
Reymonta 4, 30-059 Kraków, Poland

**Abstract**

We investigate a version of the abelian Higgs model with a non-standard kinetic term ($K$ field theory) in 2+1 dimensions. The existence of vortex type solutions with compact support (topological compactons) is established by a combination of analytical and numerical methods. This result demonstrates that the concept of compact solitons in $K$ field theories can be extended to higher dimensions.

[*]adam@fpaxp1.usc.es
[**]klimas@fpaxp1.usc.es
[†]joaquin@fpaxp1.usc.es
[††]wereszczynski@th.if.uj.edu.pl

# 1 Introduction

Theories with functions of gradients other than quadratic have been studied already for a long time, starting with the well-known Born–Infeld theory many decades ago. Higher powers of the derivatives may be introduced in order to avoid the Derrick scaling argument, opening the possibility of static finite energy solutions (solitons), as is the case, e.g., in the Skyrme model and its generalizations [1] - [6]. These theories found some applications in strong interaction physics. Another field of applications of theories with higher kinetic terms ($K$ field theories) is cosmology, where the proposal goes under the name of $K$ essence [7] - [10]. There, the $K$ fields may influence in a nontrivial way both the global expansion of the universe and the propagation of small perturbations relevant for the matter distribution. There exists even a recent proposal combining both Skyrme and $K$ essence concepts ([11]).

One interesting contribution was the observation [12] that a quartic kinetic term can produce defects with compact support (or compactons [13] - [16]) with typical regular potentials. This idea, restricted to one dimension and scalar fields, lead to a rather natural application in brane cosmology, see [17] - [20], to which we refer also for a more detailed introduction and reference list. The extension to higher dimensions and gauge fields is therefore of interest.

Let us describe now in more detail the theory which we want to study. We shall be concerned with a specific class of topological defects which may form in theories with a non-standard kinetic term (K field theories), namely topological defects with a compact support (compactons). Compactons have been mainly investigated in the form of compact topological solitons in 1+1 dimensional field theories (or, equivalently as compact domain walls with dimension $d-1$ in $d+1$ dimensional field theories that is, with co-dimension 1 in $d$ space dimensions). Compact solitons in 1+1 dimensional field theories may form for different reasons. One possibility is a potential for the field which has a non-continuous first derivatve at (some of) its vacuum values, a so-called V-shaped (or W-shaped) potential. Another possibility consists in a non-standard kinetic term (K term) with a certain behaviour at low energies (absence of the normal, quadratic kinetic term in the limit of low energy). In this latter case, the potential term should still possess more than one vacuum (in order to allow for topological solitons), but may be of the standard U shape otherwise. It is the purpose of the present paper to generalize the investigation of compact topological defects for K field theories to higher dimensions or, equivalently, to compact topological defects with a co-dimension greater than one.

As for conventional solitons, also in the case of compact K field solitons there are some significant differences between solitons in one space dimension on the one hand, and solitons in more than one space dimension, on the other hand. So let us first briefly describe some results of compact K field solitons in 1+1 spacetime dimensions. Concretely, in [12] a Lagrangian density

$$L = \tilde{M}^2 |\xi_\mu \xi^\mu| \xi_\mu \xi^\mu - 3\lambda^2 (\xi^2 - a^2)^2, \qquad (1)$$

was introduced, where $\xi$ is a real scalar field, and $\xi_\mu \equiv \partial_\mu \xi$, etc. Here the potential term is just the standard quartic potential with the two vacuum values $\xi = \pm a$, whereas the kinetic term is nontrivial (quartic in this specific example). Further, a Minkowski metric is assumed, that is, $\xi^\mu \xi_\mu = \xi_t^2 - \xi_x^2$, etc. The static field equation resulting from the above Lagrangian is

$$\xi_x^2 \xi_{xx} - \frac{\lambda^2}{\tilde{M}^2}(\xi^2 - a^2)\xi = 0. \qquad (2)$$

A compacton is defined by the condition that the field $\xi$ approaches its vacuum value $\xi = \pm a$ for finite $x$. The above static field equation is fulfilled provided that the spatial gradient term $\xi_x^2$ is zero, as well, which is true for a constant $\xi = \pm a$. At this point the difference with compact topological solitons in higher dimensions is quite obvious. In fact, for higher-dimensional topological compact solitons the vacuum manifold of the potential will no longer be a discrete set of vacuum values. Instead, it will be a circle (for vortex type compactons), a two-sphere (for monopole type compactons), etc. Correspondingly, the compacton field will no longer be a single scalar field, but rather a complex scalar, a three component field taking values in the adjoint of SU(2), etc. Further, a topologically nontrivial compacton is defined by the condition that the compacton field approaches the vacuum manifold at a finite radius such that it takes values in the full vacuum manifold (that is, covers the full circle, two-sphere, etc.). But this behaviour is not compatible with a vanishing spatial gradient term $(\nabla \xi)^2 = 0$, because the angular parts of the gradient are necessarily nonzero. This whole reasoning is, in fact, quite similar to the argument which demonstrates the non-existence of topological solitons in higher-dimensional scalar field theories, and also the way out is the same. It consists in the introduction of gauge fields, such that the ordinary gradient is replaced by a covariant gradient, and the gauge field may exactly compensate for the nonzero angular gradient (that is, the scalar field taking values in the vacuum manifold is, in fact, a pure gauge configuration).

Concretely, we shall focus on the case of a complex scalar field coupled to an abelian gauge field in 2+1 space-time dimensions, which will give rise to compactons of the vortex type. The general discussion above applies equally well to a scalar field in the adjoint representation of SU(2), coupled to a nonabelian SU(2) gauge field, where compactons of the monopole type should exist. It turns out, however, that the case of monopole type compactons is technically and calculationally much more involved, and not all results which are found in the vortex compacton case can be achieved for the monopole compacton. Therefore, we restrict to the vortex case in this paper.

It is a generic feature of compacton field configurations that they are continuous with a continuous first derivative at the compacton boundary, whereas the second derivative is discontinuous. Further, the gauge field enters into the covariant derivative in an analogous fashion as the derivative operator, therefore one could expect that the gauge field might behave like the first derivative of the compacton field at the compacton boundary, that is, continuous with a noncontinuous first derivative. This is indeed what happens if the kinetic term for the gauge field is of the standard Maxwell form. This implies that the gauge field is only a weak solution to its field equation (that is, it does not solve the field equation at the compacton boundary). Further, the contribution of the Maxwell term to the energy density is discontinuous at the compacton boundary. The energy density has, however, no singularities and the resulting total energy is finite, rendering these solutions acceptable from the point of view of a finite energy condition. Whether these weak solutions are acceptable physically depends on the physical system or physical problem under investigation. There exists the possibility to have gauge fields with a continuous first derivative and discontinuous second derivative, like the compacton field itself. This requires, however, the introduction of a non-standard kinetic term for the gauge field, as well (instead of the standard Maxwell term). Solutions of this second type are standard (i.e., they are not weak but, instead, hold in all space), because the discontinuity of the second derivatives at the compacton boundary is always suppressed by a multiplying factor zero, due to the non-standard kinetic terms (both for the scalar field and the gauge field).

In Section 2 we investigate the case of a standard kinetic term for the gauge field. We choose the usual rotationally symmetric ansatz for the scalar and gauge fields and study the resulting system of ODEs both analytically and numerically. In Section 3 we perform the same analysis for the case of a non-standard kinetic term for the gauge field. Section 4 contains our conclusions.

## 2 Compact vortices for standard gauge field kinetic term

We study the following action

$$S = \int d^3x [K(X) - V(\phi) - \frac{1}{4}F_{\mu\nu}F^{\mu\nu}] \qquad (3)$$

where $\phi$ is a complex scalar field and the potential

$$V(\phi) = \frac{\lambda}{4}\left(|\phi|^2 - v^2\right)^2 \qquad (4)$$

is the usual mexican hat potential which takes its minimum value $V = 0$ at $|\phi| = v$. Further,

$$F_{\mu\nu} \equiv \partial_\mu A_\nu - \partial_\nu A_\mu \qquad (5)$$

is the field strength tensor of the abelian gauge potential $A_\mu$, $X$ is the modulus squared of the covariant derivative

$$X \equiv (D_\mu \phi)(D^\mu \phi)^* \quad , \qquad D_\mu \equiv \partial_\mu - ieA_\mu \qquad (6)$$

and $K$ is an (at the moment arbitrary) function of its argument (the nonstandard kinetic term). The kinetic term for the gauge field, on the other hand, is given by the standard Maxwell term in this section. Here we assume that any dimensionful constants have been absorbed by a rescaling of the coordinates and fields, such that $x$, $\phi$ and $A_\mu$ are dimensionless. Consequently, $\lambda$, $v$ and $e$ are dimensionless constants. Further, our signature for the Minkowski metric in 1+2 dimensions is $(+,-,-)$. We remark that models of this type, allowing for gauge $K$ vortices, have been studied recently in [21], although not for compacton type solutions.

The Euler–Lagrange equations resulting from this action are

$$K_X D_\mu D^\mu \phi + K_{XX} X_{,\mu} D^\mu \phi + V_{\phi^*} = 0 \qquad (7)$$
$$\partial_\mu F^{\mu\nu} = ej^\nu \qquad (8)$$

(here $K_X \equiv \frac{dK}{dX}$ etc.), where the (conserved) current $j_\mu$ is

$$j_\mu = -iK_X[\phi^* D_\mu \phi - \phi(D_\mu \phi)^*]. \qquad (9)$$

As we want to study static vortex solutions, we choose the ansatz for the simplest vortex with winding number one,

$$\phi(x) = e^{i\varphi} f(r) \qquad (10)$$
$$A_j(x) = -\frac{1}{e}\frac{\alpha(r)}{r^2}\epsilon_{jk}x^k \qquad (11)$$

as well as $A_0 = 0$, where $j, k = 1, 2$, $r$ and $\varphi$ are polar coordinates

$$x^1 = r\cos\varphi \quad , \quad x^2 = r\sin\varphi \tag{12}$$

and $f(r)$ and $\alpha(r)$ are at the moment arbitrary functions of their argument. With this ansatz, for $X$ we get

$$X = -\left(f'^2 + \frac{(1-\alpha)^2}{r^2}f^2\right) \tag{13}$$

where the prime denotes derivative w.r.t. $r$. Then, the resulting equations for $f$ and $\alpha$ are

$$-K_X\left(f'' + \frac{1}{r}f' - \frac{(1-\alpha)^2}{r^2}f\right) - K_{XX}X'f' + \frac{\lambda}{2}(f^2 - v^2)f = 0, \tag{14}$$

$$\left(\frac{\alpha'}{r}\right)' + 2\frac{e^2}{r}f^2(1-\alpha)K_X = 0. \tag{15}$$

Next, we want to make a specific choice for the nonstandard kinetic term $K$. Concretely, we choose

$$K = \frac{1}{2}|X|X \tag{16}$$

which for static configurations is equal to

$$K = -\frac{1}{2}X^2 \tag{17}$$

Remark: as far as static configurations are concerned, we could choose the kinetic term $K = -\frac{1}{2}X^2$ from the very beginning. However, for time dependent configurations this kinetic term in general does not lead to an energy bounded from below, whereas the expression $K = \frac{1}{2}|X|X$ does lead to a bounded energy.

With this choice for the kinetic term we get the field equations

$$\left(f'^2 + \frac{(1-\alpha)^2}{r^2}f^2\right)\left(f'' + \frac{1}{r}f' - \frac{(1-\alpha)^2}{r^2}f\right) + \left(f'^2 + \frac{(1-\alpha)^2}{r^2}f^2\right)' f'$$

$$- \frac{\lambda}{2}(f^2 - v^2)f = 0 \tag{18}$$

and

$$\left(\frac{\alpha'}{r}\right)' + 2\frac{e^2}{r}f^2(1-\alpha)\left(f'^2 + \frac{(1-\alpha)^2}{r^2}f^2\right) = 0. \tag{19}$$

We observe that for our specific choice for the kinetic term, the constant $v$ may be brought to the value of $v = 1$ by a dimensionless rescaling of $f$ (that is, of $\phi$). Indeed, as both the kinetic term and the potential are quartic in $\phi$, the field equation for $\phi$ (or $f$) is of the third power and, therefore, homogeneous. In the Maxwell equation, on the other hand, the rescaling of $\phi$ results in a rescaling of the current $j_\mu$ which is, again, homogeneous in $\phi$. This rescaling may be compensated by a redefinition of the (dimensionless) electric charge $e$. Therefore, we may set $v = 1$ in the above system of equations without loss of generality, which we assume in the sequel.

## 2.1 Expansion about the center

We now want to insert a power series expansion about the center $r = 0$ into the above equations. It is easy to find that only odd powers contribute to $f$, whereas only even powers contribute for $\alpha$,

$$f(r) = \sum_{n=1}^{\infty} A_{2n-1} r^{2n-1} \tag{20}$$

$$\alpha(r) = \sum_{n=1}^{\infty} a_{2n} r^{2n} \tag{21}$$

Here $a_2$ and $A_1$ are free parameters, whereas the higher coefficients are determined in terms of $a_2$, $A_1$, $\lambda$ and $e$. Introducing the notation

$$a \equiv a_2 \quad , \quad A \equiv A_1 \tag{22}$$

we get concretely for the first few coefficients

$$a_4 = -\frac{1}{2} e^2 A^4 \tag{23}$$

$$a_6 = \frac{1}{64} e^2 \lambda A^2 + \frac{1}{3} e^2 A^4 a \tag{24}$$

$$A_3 = -\frac{1}{64} \frac{\lambda}{A} \tag{25}$$

$$A_5 = -\frac{1}{49152} \frac{1}{A^3} [1024 A^4 (e^2 A^4 - a^2) + 192 \lambda A^2 a - 256 \lambda A^4 + 15 \lambda^2] \tag{26}$$

and one finds that due to the nonlinearity of the system the higher coefficients are quite complicated.

## 2.2 Expansion about the boundary

Now we assume that there exists a compacton boundary, that is, a value $r = R$ such that the field $f$ approaches its vacuum value, and that the first derivative is zero,

$$f(r = R) = 1 \quad , \quad f'(r = R) = 0. \tag{27}$$

Further, we assume that $\alpha$ takes its vacuum value at the same point $r = R$, $\alpha(r = R) = 1$. We will see in a moment that we cannot assume $\alpha'(r = R) = 0$ if we want to get nontrivial results.

We remark that the local analysis of this subsection cannot be used to determine the value of $R$ where the fields approach their vacuum values. This value can be determined either by a complete analytic solution of the system (which is out of reach in the present case), or by a numerical integration, starting with the conditions determined by the local analysis (the power series expansions) at one boundary (e.g. $r = 0$), and using a shooting algorithm to reach the other boundary (e.g. $r = R$), with the boundary conditions again given by the local analysis. This numerical integration shall be performed in the next subsection.

It is useful to introduce the new variable

$$\epsilon = R - r \tag{28}$$

and to subtract the vacuum values of the fields, that is

$$f(r) \equiv 1 - g(\epsilon) \quad , \quad \alpha(r) \equiv 1 - \beta(\epsilon) \tag{29}$$

With this change we obtain the following system of equations

$$\left( g'^2 + \frac{\beta^2}{(R-\epsilon)^2}(1-g)^2 \right) \left( -g'' + \frac{g'}{R-\epsilon} + \frac{\beta^2}{(R-\epsilon)^2}(1-g) \right) -$$

$$\left( g'^2 + \frac{\beta^2}{(R-\epsilon)^2}(1-g)^2 \right)' g' + \frac{1}{2}\lambda g(1-g)(2-g) = 0 \tag{30}$$

and

$$-\beta'' - \frac{\beta'}{R-\epsilon} + 2e^2(1-g)^2\beta \left( g'^2 + \frac{\beta^2}{(R-\epsilon)^2}(1-g)^2 \right) = 0 \tag{31}$$

where now the prime denotes derivative w.r.t. $\epsilon$. Now we insert the power series expansion

$$g(\epsilon) = \sum_{n=2}^{\infty} B_n \epsilon^n \tag{32}$$

$$\beta(\epsilon) = \sum_{n=1}^{\infty} b_n \epsilon^n \qquad (33)$$

into the above equations. There is only one free parameter, namely $b_1$, because we have fixed the three conditions

$$g(\epsilon = 0) = 0 \quad g'(\epsilon = 0) = 0\,, \quad \beta(\epsilon = 0) = 0. \qquad (34)$$

Introducing the notation

$$b \equiv b_1 \qquad (35)$$

we find for the coefficient $B_2$ a cubic equation with the three solutions

$$B_2 = 0, \pm \frac{1}{12R}\sqrt{6\lambda R^2 - 36b^2}. \qquad (36)$$

This implies that we may indeed join the vacuum solution $B_2 = 0$ with the compacton solution (the positive root in the above solution) at the compacton boundary $\epsilon = 0$. For the compacton we choose

$$B_2 = +\frac{1}{12R}\sqrt{6\lambda R^2 - 36b^2} \qquad (37)$$

which implies the inequality

$$6b^2 \leq \lambda R^2. \qquad (38)$$

The other coefficients for the compacton are determined uniquely by linear equations and the first few are given by

$$B_3 = \frac{1}{36R^2} \frac{\lambda R^2 - 24b^2}{5\lambda R^2 - 24b^2} \sqrt{6\lambda R^2 - 36b^2} \qquad (39)$$

$$b_2 = -\frac{b}{2R}, \quad b_3 = 0\,, \quad b_4 = 0 \qquad (40)$$

$$b_5 = \frac{e^2}{60}\lambda b\,, \quad b_6 = -\frac{e^2 \lambda b}{360R} \frac{11\lambda R^2 - 72b^2}{5\lambda R^2 - 24b^2} \qquad (41)$$

We do not display higher coefficients, because already the expression for the coefficient $B_4$ is a rather complicated three-line expression. It is obvious from the above expressions that the linear coefficient $b \equiv b_1$ of the gauge field must be nonzero, because for $b = 0$ all the higher $b_i$ are zero, as well, leading to a gauge field which is pure gauge in the whole space $\mathbb{R}^2 \setminus \{\mathbf{0}\}$, which cannot provide a finite energy solution, as discussed already in the Introduction.

## 2.3 Numerical evaluation

There are two possibilities for a numerical integration of our system. We may either use the original system of equations (18), (19) (with $v = 1$) and start the integration at the center $r = 0$ with the initial conditions deteremined in Subsection 2.1 (power series expansion at the center). Then we require that there exists a radius $r = R$ such that at this point the numerical solution obeys the boundary conditions determined in Subsection 2.2 (expansion at the boundary). This procedure we call shooting from the center. In this case we have three free parameters at our disposal, namely $a, A$ and $R$. At the same time, we have to fulfill three conditions at the boundary, namely $f(R) = 1, f'(R) = 0$ and $\alpha(R) = 1$. Therefore, we expect a solution to exist in the generic case, that is, for arbitrary values of the two coupling constants of the theory (the electric charge $e$ and the strength of the Higgs potential, $\lambda$).

The same conclusion can be reached by analysing the shooting from the boundary, instead. In this case, we use the system of equations (30), (31) for the numerical integration. Further, we have two free parameters in this case, namely $b$ (that is, $\alpha'(R)$), and $R$. At the same time, we have two conditions to obey at the center, namely $f(r = 0) = 0$ and $\alpha(r = 0) = 0$. Therefore, the number of adjustable free parameters again matches the number of conditions, and we expect that a solution will exist generically. We remark that the condition $\alpha'(r = 0) = 0$ does not count as an additional boundary condition, because it is a consequence of the symmetries of the equations of motion (that is, $\alpha(r)$ has a power series expansion about $r = 0$ in terms of $r^2$ rather than $r$).

Concretely, we use the shooting from the boundary for the numerical integration, because it is numerically simpler (there are only two free adjustable parameters and two boundary conditions). In Figures 1-4 we show the result of the numerical integration for some selected values of the coupling constants $e$ and $\lambda$. We see that the behaviour determined from the power series expansions in Sections 2.1 and 2.2 is exactly reproduced by the numerical solutions.

## 3 Compact vortices for non-standard gauge field kinetic term

Now we study the action

$$S = \int d^3x [K(X) - V(\phi) - \mathcal{F}^n] \qquad (42)$$

where
$$\mathcal{F} \equiv \frac{1}{4}F_{\mu\nu}F^{\mu\nu} \tag{43}$$

and $n$ is an integer whose value will be determined in a moment. So in this section the kinetic term of the gauge field is non-standard, as well. The remaining terms are as in Section 2. The Euler–Lagrange equation for the scalar field $\phi$ is identical to the one in Section 2, whereas for the gauge field it is
$$\partial_\mu(n\mathcal{F}^{n-1}F^{\mu\nu}) = ej^\nu \tag{44}$$

and the current is again
$$j^\nu = -K_X[\phi^* D^\nu \phi - \phi(D^\nu \phi)^*] \tag{45}$$

Using again the radially symmetric ansatz (10), (11), we get
$$\mathcal{F} = \frac{1}{2}\left(\frac{\alpha'}{r}\right)^2 \tag{46}$$

and, for Eq. (44) we find
$$\frac{(2n-1)n}{2^{n-1}}\left(\frac{\alpha'}{r}\right)^{2n-2}\left(\frac{\alpha'}{r}\right)' + 2e^{2n}K_X\frac{f^2}{r}(1-\alpha) = 0. \tag{47}$$

For the specific choice $K = -\frac{1}{2}X^2$ for static configurations, we get
$$\frac{(2n-1)n}{2^{n-1}}\left(\frac{\alpha'}{r}\right)^{2n-2}\left(\frac{\alpha'}{r}\right)' + 2e^{2n}\left(f'^2 + \frac{(1-\alpha)^2}{r^2}f^2\right)\frac{f^2}{r}(1-\alpha) = 0. \tag{48}$$

## 3.1 Expansion at the boundary

We first perform the expansion at the boundary, because this will serve to determine the value $n$ of the integer power of the gauge field kinetic term. We, again, introduce the variable $\epsilon \equiv R - r$ and the functions $g(\epsilon)$ and $\beta(\epsilon)$, like in (28), (29). Then the resulting field equations in the variable $\epsilon$ are (30) for the scalar field and
$$P_n + Q_n = 0 \tag{49}$$

for the gauge field, where
$$P_n \equiv \frac{(2n-1)n}{2^{n-1}}\left(\frac{\beta'}{R-\epsilon}\right)^{2n-2}\left(-\frac{\beta''}{R-\epsilon} - \frac{\beta'}{(R-\epsilon)^2}\right) \tag{50}$$

and
$$Q_n \equiv 2e^{2n}\frac{(1-g)^2}{R-\epsilon}\left(g'^2 + \frac{\beta^2(1-g)^2}{(R-\epsilon)^2}\right)\beta. \tag{51}$$

Next, we introduce the expansions about the boundary

$$g(\epsilon) = \sum_{k=2} \tilde{B}_k \epsilon^k \tag{52}$$

and

$$\beta(\epsilon) = \sum_{k=2} \tilde{b}_k \epsilon^k \tag{53}$$

where we assume that both $g$ and $\beta$ start with the quadratic term, that is, a quadratic approach to the vacuum value. Inserting now these expansions into Eq. (49), we find in leading order

$$P_n = -(2n-1)n2^n\left(\frac{\tilde{b}_2}{R}\right)^{2n-1}\epsilon^{2n-2} + \mathcal{O}(\epsilon^{2n-1}) \tag{54}$$

$$Q_n = e^{2n}\frac{8\tilde{b}_2\tilde{B}_2^2}{R}\epsilon^4 + \mathcal{O}(\epsilon^5) \tag{55}$$

and, therefore, a necessary condition for the cancellation of the leading order is $2n - 2 = 4$ or

$$n = 3 \tag{56}$$

which we assume in the sequel. Cancellation of the leading order now leads to the condition

$$\tilde{b}_2(e^6 R^4 \tilde{B}_2^2 - 15\tilde{b}_2^4) = 0 \tag{57}$$

with the five solutions

$$\tilde{b}_2 = 0, \pm e^{\frac{3}{2}}R\left(\frac{\tilde{B}_2^2}{15}\right)^{\frac{1}{4}}, \pm ie^{\frac{3}{2}}R\left(\frac{\tilde{B}_2^2}{15}\right)^{\frac{1}{4}} \tag{58}$$

The solution $\tilde{b}_2 = 0$ corresponds to the vacuum solution, whereas the positive, real solution corresponds to the compacton. The remaining coefficients are determined by inserting the power series expansion (52), (53) into the system of equations (30) and (49). The leading order coefficient $\tilde{B}_2$ is, in fact, equal to the leading order coefficient $B_2$ of Subsection 2.2 (but for the value $b \equiv b_1 = 0$), because equation (30) is the same in both cases. Therefore, we get

$$\tilde{B}_2 = 0, \pm\frac{\sqrt{6\lambda}}{12} \tag{59}$$

and the compacton corresponds to the choice $B_2 = \frac{\sqrt{6\lambda}}{12}$. For the compacton value for $\tilde{b}_2$ we get correspondingly

$$\tilde{b}_2 = R\left(\frac{e^6\lambda}{360}\right)^{\frac{1}{4}} \tag{60}$$

and the higher coefficients are uniquely determined by linear equations. We find, for instance,

$$\tilde{B}_3 = \frac{\sqrt{6\lambda}}{180R} \quad , \quad \tilde{B}_4 = -\frac{\sqrt{15}}{324}e^3 + \frac{23\sqrt{6}}{6480}\frac{\sqrt{\lambda}}{R^2} - \frac{\lambda}{144} \tag{61}$$

$$\tilde{b}_3 = \frac{3}{5}\left(\frac{e^6\lambda}{360}\right)^{\frac{1}{4}} \tag{62}$$

and

$$\tilde{b}_4 = -\frac{97}{7020}R^{-1}\left(\frac{e^6\lambda}{360}\right)^{\frac{1}{4}} + \frac{14}{117}\frac{R}{\lambda}\left(\frac{e^6\lambda}{360}\right)^{\frac{3}{4}} - \frac{2}{117}\sqrt{6\lambda}R\left(\frac{e^6\lambda}{360}\right)^{\frac{1}{4}}. \tag{63}$$

The expressions for the higher coefficients are too long to be displayed here. We remark that due to the boundary conditions imposed there are no free parameters in the expansion at the boundary. Indeed, all expansion coefficients are determined in terms of the parameters of the theory, $e$ and $\lambda$, and in terms of the compacton radius $R$.

## 3.2 Expansion at the center

We insert the power series expansions at the center

$$f(r) = \sum_{n=1}^{\infty} \tilde{A}_{2n-1} r^{2n-1} \tag{64}$$

$$\alpha(r) = \sum_{n=1}^{\infty} \tilde{a}_{2n} r^{2n} \tag{65}$$

into the system of equations (18), (48), where we set $n = 3$ in the latter equation. Analogously to Subsection 2.1., only odd powers contribute to $f$, and only even powers contribute to $\alpha$. Here, $\tilde{A}_1 \equiv \tilde{A}$ and $\tilde{a}_2 \equiv \tilde{a}$ are free parameters, and the higher coefficients can be expresssed by them. They are, in fact, uniquely determined by linear equations. Explicitly, we find

$$\tilde{a}_4 = -\frac{1}{120}\frac{e^6 \tilde{A}^4}{\tilde{a}^4} \tag{66}$$

$$\tilde{a}_6 = \frac{1}{5400}\frac{e^6\tilde{A}^4(-e^6\tilde{A}^4 + 30\tilde{a}^6)}{\tilde{a}^9} + \frac{1}{3840}\frac{e^6\tilde{A}^2}{\tilde{a}^4}\lambda \tag{67}$$

$$\tilde{A}_3 = -\frac{1}{64}\frac{\lambda}{\tilde{A}} \tag{68}$$

$$\tilde{A}_5 = \frac{1}{2880}\frac{\tilde{A}(-e^6\tilde{A}^4 + 60\tilde{a}^6)}{\tilde{a}^4} - \frac{1}{768}\frac{3\tilde{a} - 4\tilde{A}^2}{\tilde{A}}\lambda - \frac{5}{16384}\frac{\lambda^2}{\tilde{A}^3} \tag{69}$$

Again, we do not display higher coefficients.

### 3.3 Numerical evaluation

Again, there are two possibilities for a numerical intergation of our system, namely a shooting from the center or a shooting from the boundary. In both cases we will find that there exists one condition more than there exist free parameters, therefore a solution will not exist in the generic case. Instead, a fine-tuning of the two remaining free coupling constants in the lagrangian, $e$ and $\lambda$, is necessary. Differently said, we shall promote one of the two coupling constants to an additional adjustable parameter. Concretely, we will assume that the electrical charge $e$ is a given, arbitrary coupling constant, whereas $\lambda$ will be treated as an additional adjustable parameter.

With this assumption, the number of adjustable parameters and the number of conditions again match. Indeed, in the case of the shooting from the center the free parameters are $a, A, R$ and $\lambda$, and the conditions are $f(r = R) = 1$, $f'(r = R) = 0$, $\alpha(r = R) = 1$ and $\alpha'(r = R) = 0$. In the case of the shooting from the boundary, the free parameters are $R$ and $\lambda$, and the conditions are $f(r = 0) = 0$ and $\alpha(r = 0) = 0$. In both cases, the free parameters match the boundary conditions, so that we expect a solution to exist, where now $\lambda$ no longer is an independent coupling constant but, instead, has an adjusted, fixed value for a given choice of the electric charge $e$. We remark that, again, the condition $\alpha'(r = 0) = 0$ does not count as an additional boundary condition, because it is a consequence of the symmetries of the equations of motion (that is, $\alpha(r)$ has a power series expansion about $r = 0$ in terms of $r^2$ rather than $r$).

In the numerical calculations, the features described above are reproduced with a high precision. In Figure 5, we display the adjusted values of $\lambda$, for some selected values of $e$, such that a compacton of the type described in Section 3 exists. In Figure 6, we display the corresponding values of the compacton radius $R$, again as a function of $e$. It is clearly seen that the compacton radius diverges as $e \to 0$, which is as expected, because there should exist no topological compacton for zero coupling to the gauge field. In Figures 7 - 12 we plot the functions $f(r)$ and $\alpha(r)$ for some selected values of $e$. In these figures, the numerical integration is performed via shooting from the boundary, which is simpler numerically, because there are less free adjustable parameters and less boundary conditions. In all cases the figures clearly display the behaviour described in Subsections 3.1 and 3.2.

For reasons of consistency, we also performed some numerical integrations via shooting from the center. The resulting solutions $f$ and $\alpha$ are in complete agreement with the ones obtained by shooting from the boundary. As said, these calculations are more involved, because they require the determination of the correct values in a four-parameter space, whereas the shooting from the boundary only requires the determination of two parameters. Therefore, we re-calculated only some cases via shooting from the center, and we do not display the corresponding figures.

# 4 Conclusions

In this paper we investigated a system of a complex scalar field coupled to an abelian gauge field via a non-standard covariant kinetic term and established the existence of compact gauge vortices by a combination of analytical and numerical methods. Here we had to distinguish two cases. If the kinetic term of the gauge field is of the standard form, then the resulting compacton solutions are of the weak type, because the first derivative of the gauge field at the compacton boundary is discontinuous. The resulting compacton field configurations still give rise to a non-singular energy density and, consequently, to a finite total energy. For a specific non-standard choice of the gauge field kinetic term, on the other hand, we were able to establish the existence of compact vortex solutions in the sense of strong solutions, that is, solutions to the field equations in all space. Concretely, we had to choose the third power of the standard Maxwell action, which is dictated by the condition of a quadratic approach to the vacuum, like the scalar Higgs field itself. These solutions, however, do not exist for arbitrary values of the coupling constants of the theory but require, instead, a finetuning of these couplings. The reason for this finetuning is that the existence of the compacton solution imposes some boundary conditions, and in this case there are more conditions on the fields than there are free integration constants. This finetuning is, in fact, quite similar to a finetuning between the cosmological constant and the gravitational constant that was observed in [18] for a compact brane coupled to gravity in five dimensions. We remark that field theories with higher powers of the Maxwell action are not as exotic as on might think. They have been studied, in the nonabelian case, as effective low energy theories to better describe the vacuum and confinement of strongly interacting gauge field the-ories [22], [23]. They have also been introduced into cosmology where they apparently are more efficient in the creation of large scale magnetic fields, see [24].

Let us emphasize again that the absence of the usual quadratic kinetic term of the Higgs field, at least in the limit of low energies, is necessary for the exact compacton solutions to exist. As this is an unusual behaviour, one possible interpretation consists in considering these theories as effective or strong coupling limits of more standard theories in situations where propagation is not relevant, as is the case, for instance, for static solutions. The important point here is that the strong coupling limit may be a good approximation for the static solution (a compacton approximates an "almost compacton" soliton, where the (probably large) size of the compacton is controlled by the large coupling constants, whereas the tiny region of very fast exponential decay to the vacuum is controlled by the small coupling constant of the standard kinetic term), whereas it is a bad approximation for the propagation of small fluctuations (the propagation in the vacuum is completely supressed in the strong coupling limit). In any case, the study of the present paper is dedicated to establish the existence of compact vortex solutions for $K$ field theories, whereas the questions of physical relevance and possible applications shall be investigated elsewhere.

We conclude that the concept of compactons, that is, soliton solutions with compact support, can be extended to higher dimensions. In this paper, we investigated the case of higher-dimensional topological defects, specifically vortices. For these higher-dimensional topological defects, we found that finite energy solutions require the introduction of a gauge field, i.e., the study of a gauge theory. Recently, a rather general study of higher-dimensional non-topological solitons in $K$ field theories has been performed in [25], and it was found that such non-topological solitons may exist in higher dimensions under certain conditions. Therefore, also the existence of non-topological compactons in pure scalar field theories (without a gauge field) cannot be excluded, although in these cases the issue of stability will probably be more problematic. In any case, these questions are beyond the scope of the present article.

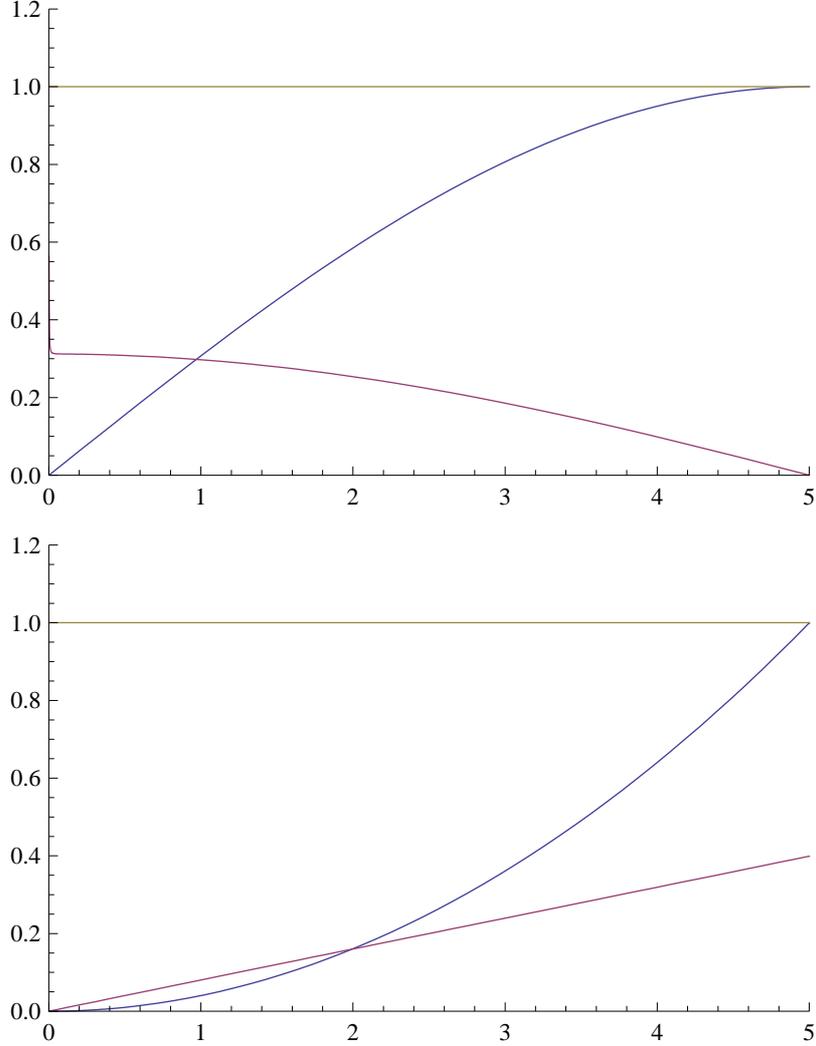

Figure 1: Standard Maxwell kinetic term, and shooting from the boundary: for $e = 0.1$ and $\lambda = 0.1$, the functions $f(r)$ and $f'(r)$ are shown in the upper diagram, whereas the functions $\alpha(r)$ and $\alpha'(r)$ are shown in the lower diagram. $f$ and $f'$ start at $f(R) = 1$ and $f'(R) = 0$ at the boundary, and are supposed to hit $f(r = 0) = 0$ and an undetermined value of $f'(r = 0)$ at the center. $\alpha$ starts at $\alpha(R) = 1$, whereas the starting value of $\alpha'(R) \equiv b$ is an adjustable parameter. $\alpha$ and $\alpha'$ are supposed to hit the values $\alpha(r = 0) = 0$ and $\alpha'(r = 0) = 0$ at the center. The adjustable parameters in this case take the values $b = 0.399135$ and $R = 5.001365$.

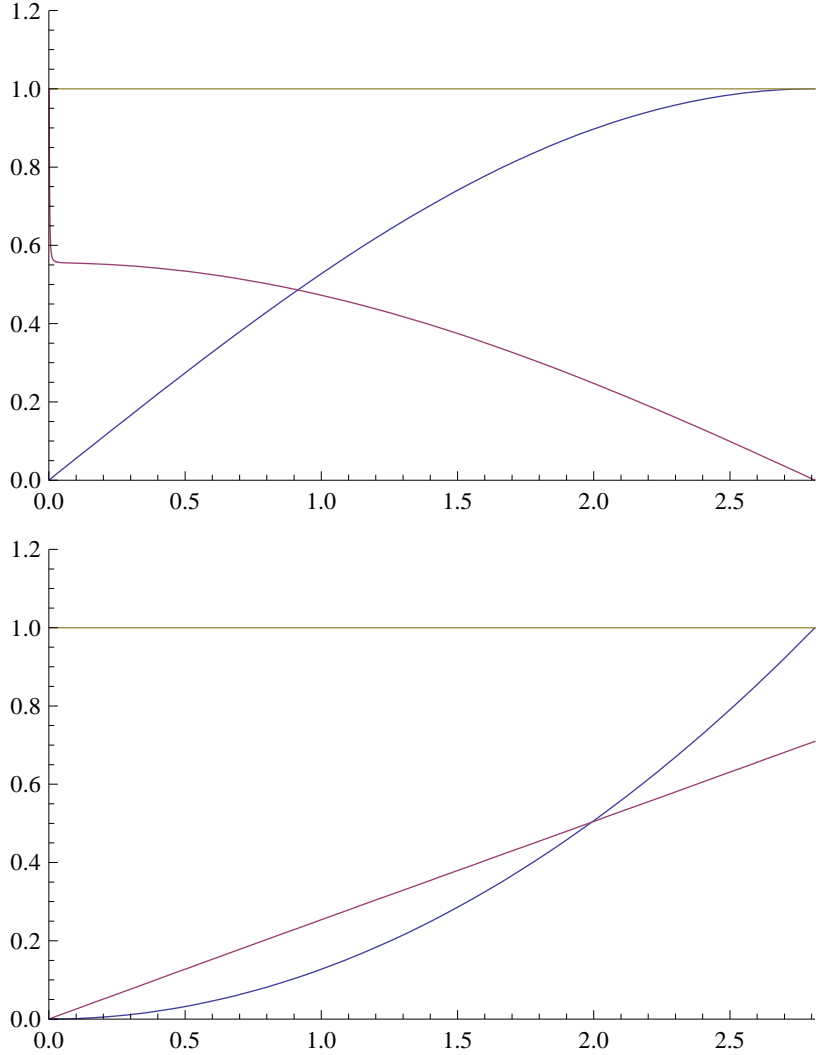

Figure 2: Standard Maxwell kinetic term, and shooting from the boundary: for $e = 0.1$ and $\lambda = 1.0$, the functions $f(r)$ and $f'(r)$ are shown in the upper diagram, whereas the functions $\alpha(r)$ and $\alpha'(r)$ are shown in the lower diagram. $f$ and $f'$ start at $f(R) = 1$ and $f'(R) = 0$ at the boundary, and are supposed to hit $f(r = 0) = 0$ and an undetermined value of $f'(r = 0)$ at the center. $\alpha$ starts at $\alpha(R) = 1$, whereas the starting value of $\alpha'(R) \equiv b$ is an adjustable parameter. $\alpha$ and $\alpha'$ are supposed to hit the values $\alpha(r = 0) = 0$ and $\alpha'(r = 0) = 0$ at the center. The adjustable parameters in this case take the values $b = 0.709773$ and $R = 2.812476$.

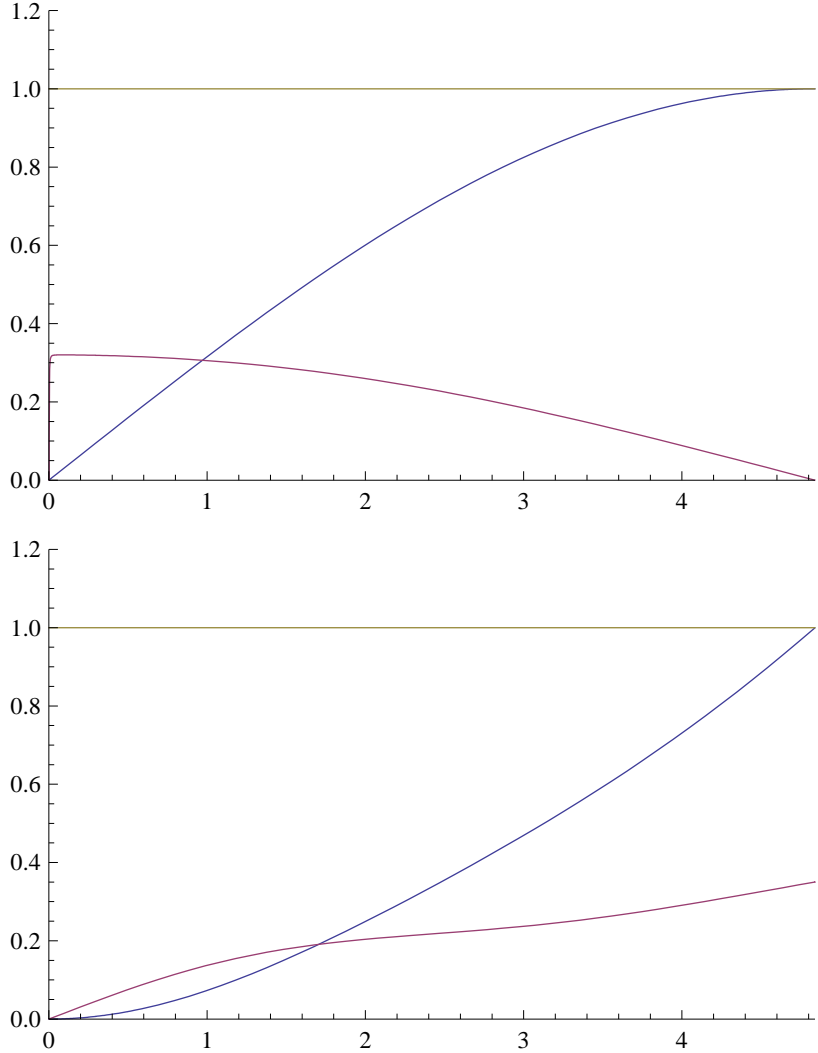

Figure 3: Standard Maxwell kinetic term, and shooting from the boundary: for $e = 1.0$ and $\lambda = 0.1$, the functions $f(r)$ and $f'(r)$ are shown in the upper diagram, whereas the functions $\alpha(r)$ and $\alpha'(r)$ are shown in the lower diagram. $f$ and $f'$ start at $f(R) = 1$ and $f'(R) = 0$ at the boundary, and are supposed to hit $f(r = 0) = 0$ and an undetermined value of $f'(r = 0)$ at the center. $\alpha$ starts at $\alpha(R) = 1$, whereas the starting value of $\alpha'(R) \equiv b$ is an adjustable parameter. $\alpha$ and $\alpha'$ are supposed to hit the values $\alpha(r = 0) = 0$ and $\alpha'(r = 0) = 0$ at the center. The adjustable parameters in this case take the values $b = 0.350292$ and $R = 4.84096$.

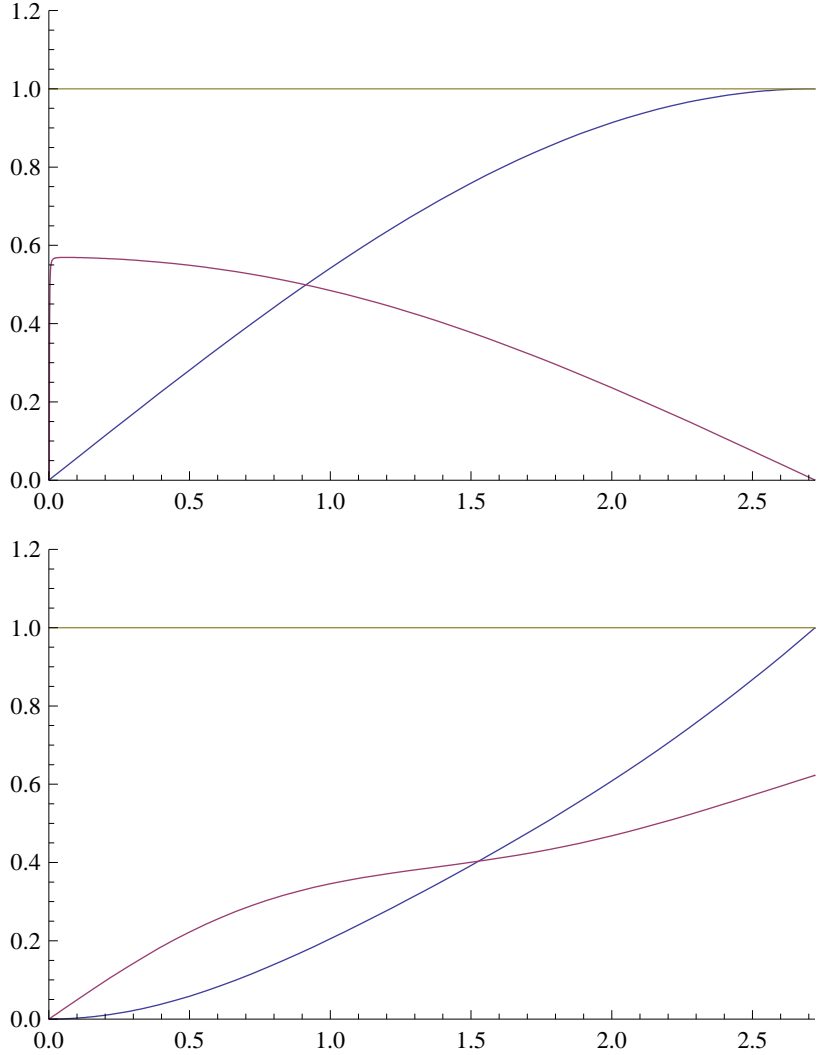

Figure 4: Standard Maxwell kinetic term, and shooting from the boundary: for $e = 1.0$ and $\lambda = 1.0$, the functions $f(r)$ and $f'(r)$ are shown in the upper diagram, whereas the functions $\alpha(r)$ and $\alpha'(r)$ are shown in the lower diagram. $f$ and $f'$ start at $f(R) = 1$ and $f'(R) = 0$ at the boundary, and are supposed to hit $f(r = 0) = 0$ and an undetermined value of $f'(r = 0)$ at the center. $\alpha$ starts at $\alpha(R) = 1$, whereas the starting value of $\alpha'(R) \equiv b$ is an adjustable parameter. $\alpha$ and $\alpha'$ are supposed to hit the values $\alpha(r = 0) = 0$ and $\alpha'(r = 0) = 0$ at the center. The adjustable parameters in this case take the values $b = 0.622917$ and $R = 2.72227$.

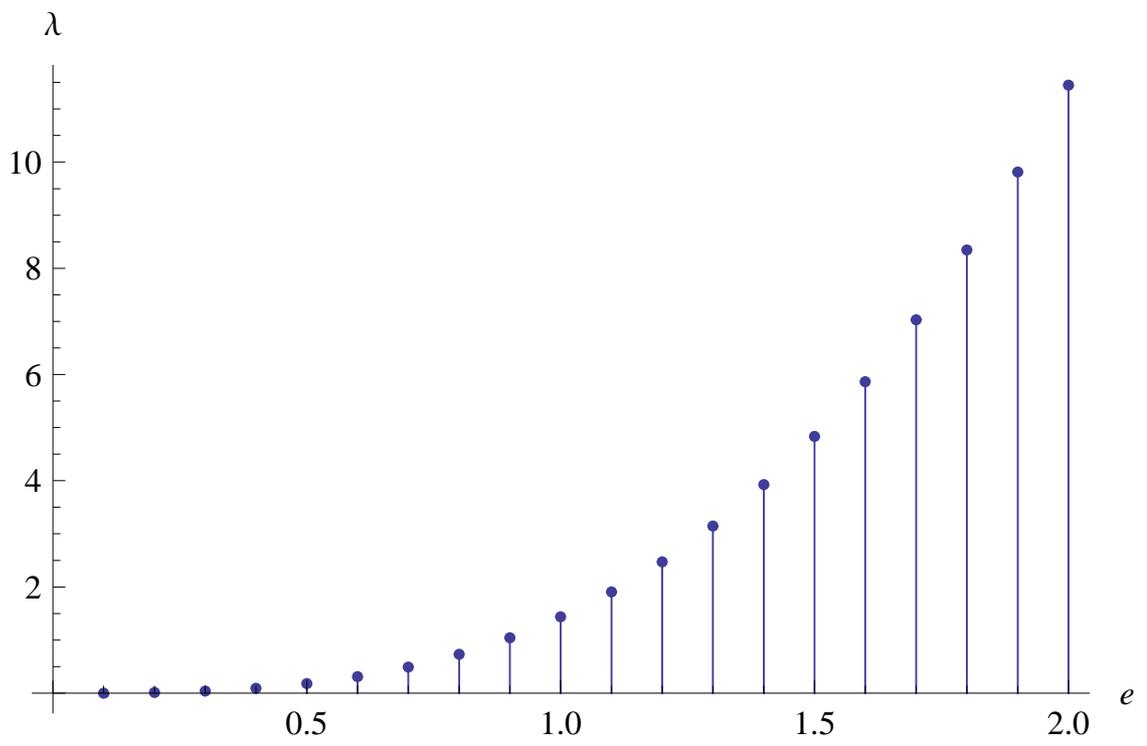

Figure 5: Non-standard gauge field kinetic term: the values of $\lambda$, for some selected values of $e$, such that a compacton exists, in the $e - \lambda$ plane.

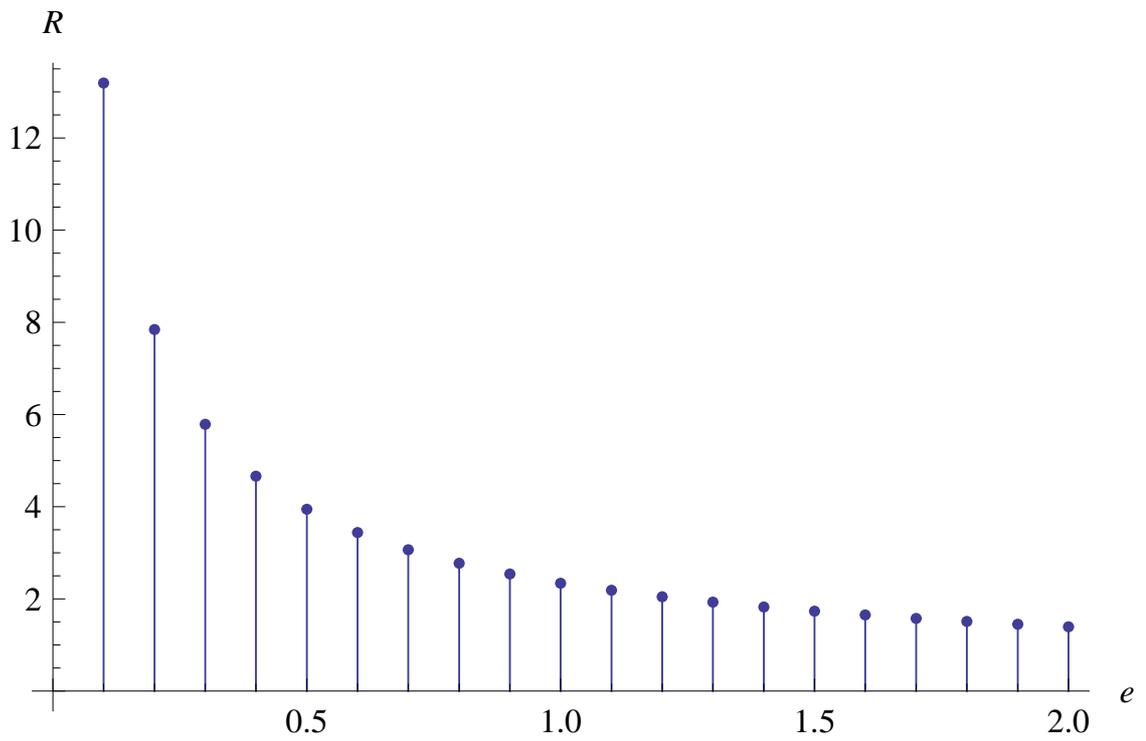

Figure 6: Non-standard gauge field kinetic term: the compacton radius $R$, for some selected values of $e$ (and for the corresponding, adjusted values of $\lambda$, such that the compacton exists).

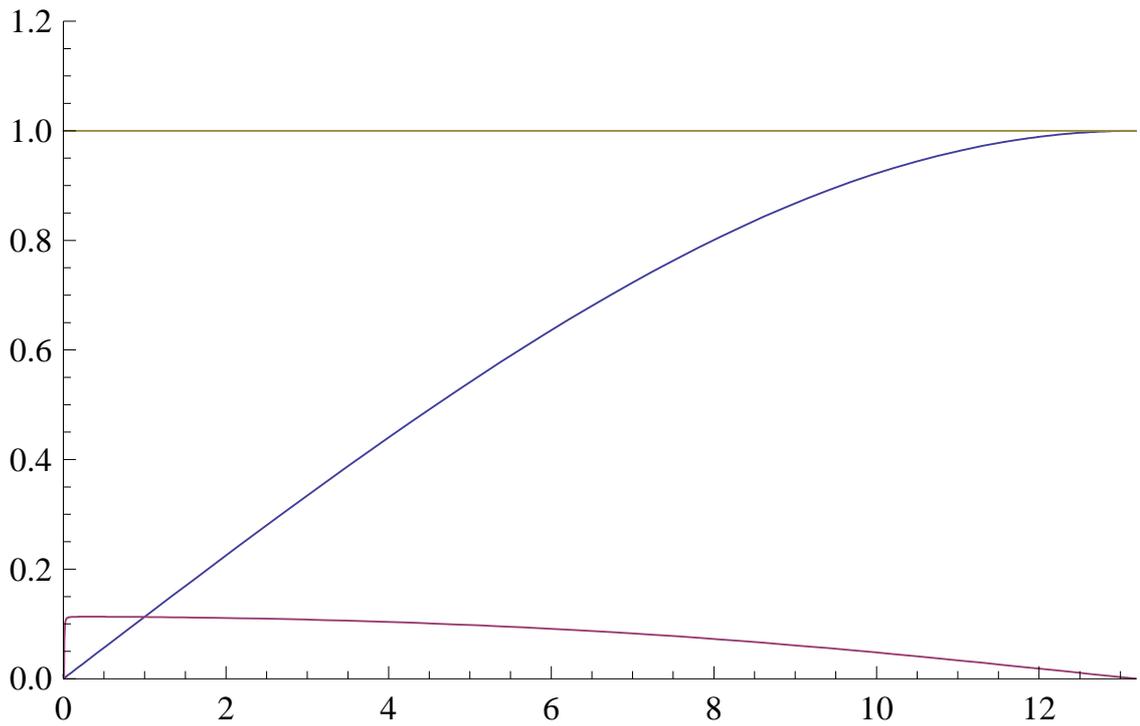

Figure 7: Non-standard gauge field kinetic term, and shooting from the boundary: for $e = 0.1$, and for the corresponding values $\lambda = 0.00143143$ and $R = 13.19195$, the functions $f(r)$ and $f'(r)$ are shown. It is clearly seen that $f$, which starts at $f(R) = 1$, goes to zero at $r = 0$, whereas $f'$, which starts at $f'(R) = 0$, goes to some nonzero value which is not determined by the asymptotic analysis.

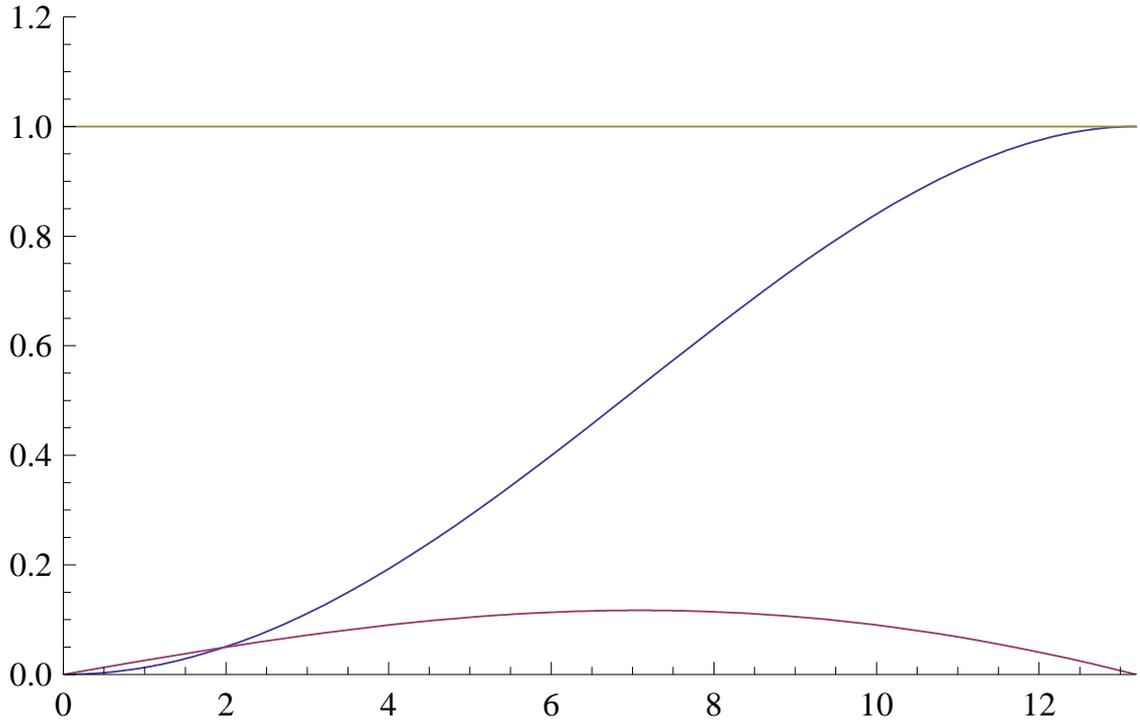

Figure 8: Non-standard gauge field kinetic term, and shooting from the boundary: for $e = 0.1$, and for the corresponding values $\lambda = 0.00143143$ and $R = 13.19195$, the functions $\alpha(r)$ and $\alpha'(r)$ are shown. It is clearly seen that $\alpha$, which starts at $\alpha(R) = 1$, goes to zero at $r = 0$, and $\alpha'$, which starts at $\alpha'(R) = 0$, goes to zero, as well. Observe that the latter condition does not count as an independent boundary condition, because it is dictated by the symmetries of the corresponding differential equation (i.e., $\alpha$ has a power series expansion about $r = 0$ in terms of $r^2$ rather than $r$).

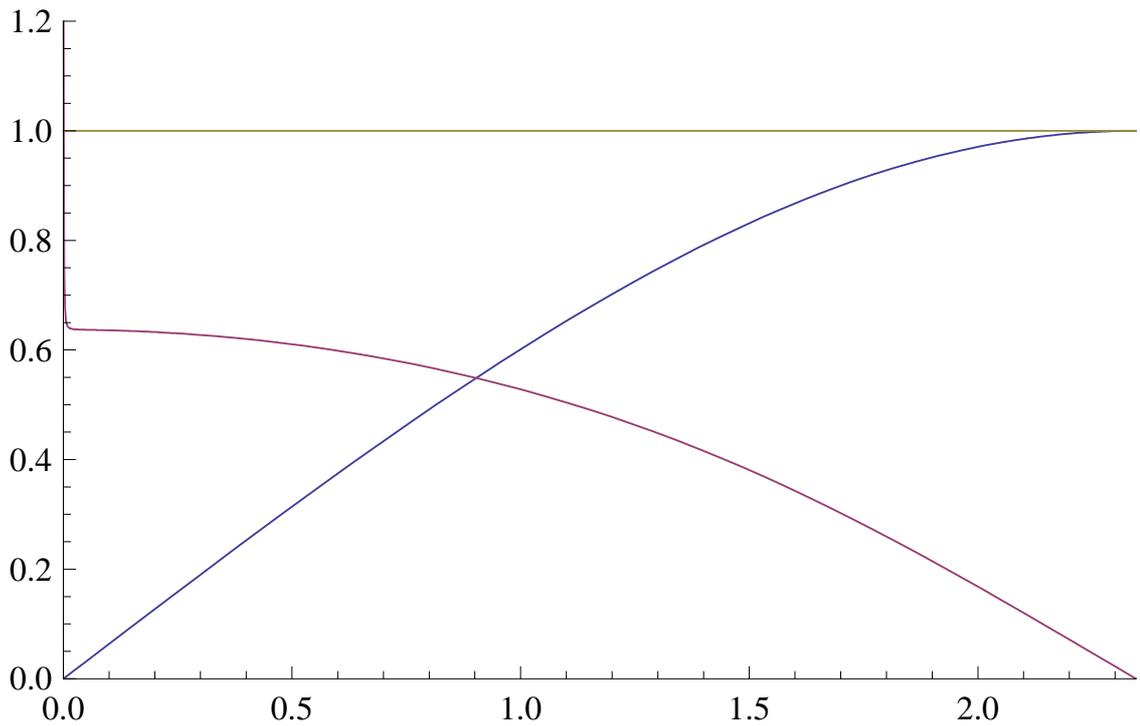

Figure 9: Non-standard gauge field kinetic term, and shooting from the boundary: for $e = 1.0$, and for the corresponding values $\lambda = 1.43144$ and $R = 2.3459$, the functions $f(r)$ and $f'(r)$ are shown. It is clearly seen that $f$, which starts at $f(R) = 1$, goes to zero at $r = 0$, whereas $f'$, which starts at $f'(R) = 0$, goes to some nonzero value which is not determined by the asymptotic analysis.

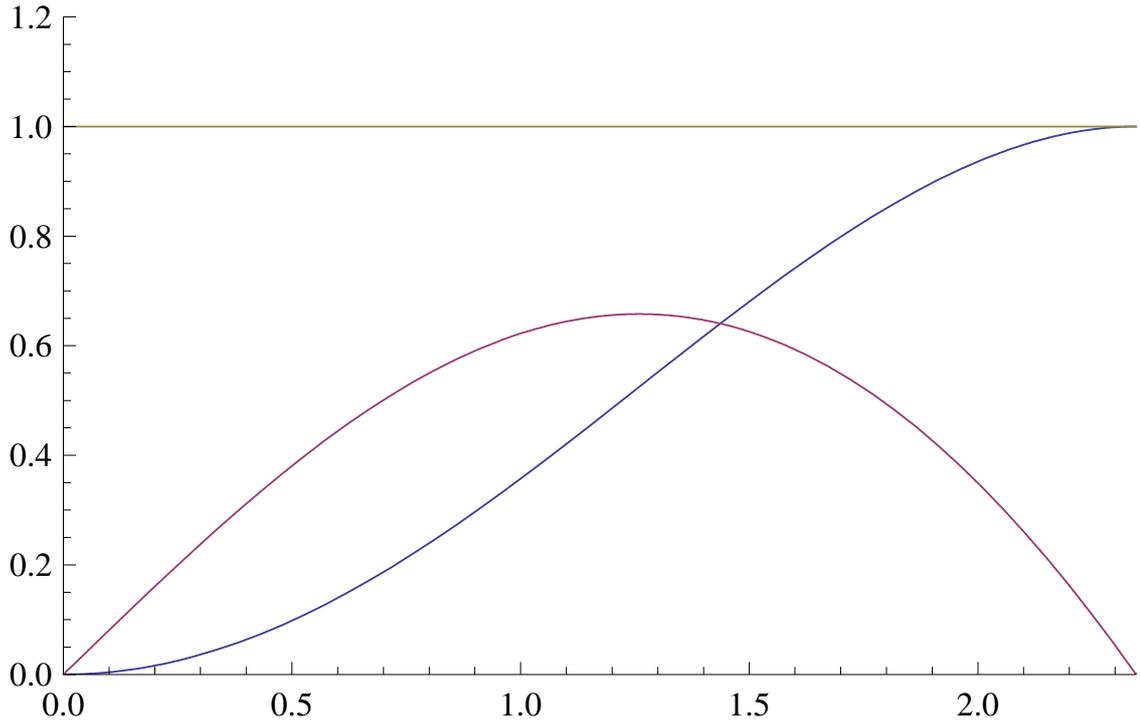

Figure 10: Non-standard gauge field kinetic term, and shooting from the boundary: for $e = 1.0$, and for the corresponding values $\lambda = 1.43144$ and $R = 2.3459$, the functions $\alpha(r)$ and $\alpha'(r)$ are shown. It is clearly seen that $\alpha$, which starts at $\alpha(R) = 1$, goes to zero at $r = 0$, and $\alpha'$, which starts at $\alpha'(R) = 0$, goes to zero, as well. Observe that the latter condition does not count as an independent boundary condition, because it is dictated by the symmetries of the corresponding differential equation (i.e., $\alpha$ has a power series expansion about $r = 0$ in terms of $r^2$ rather than $r$).

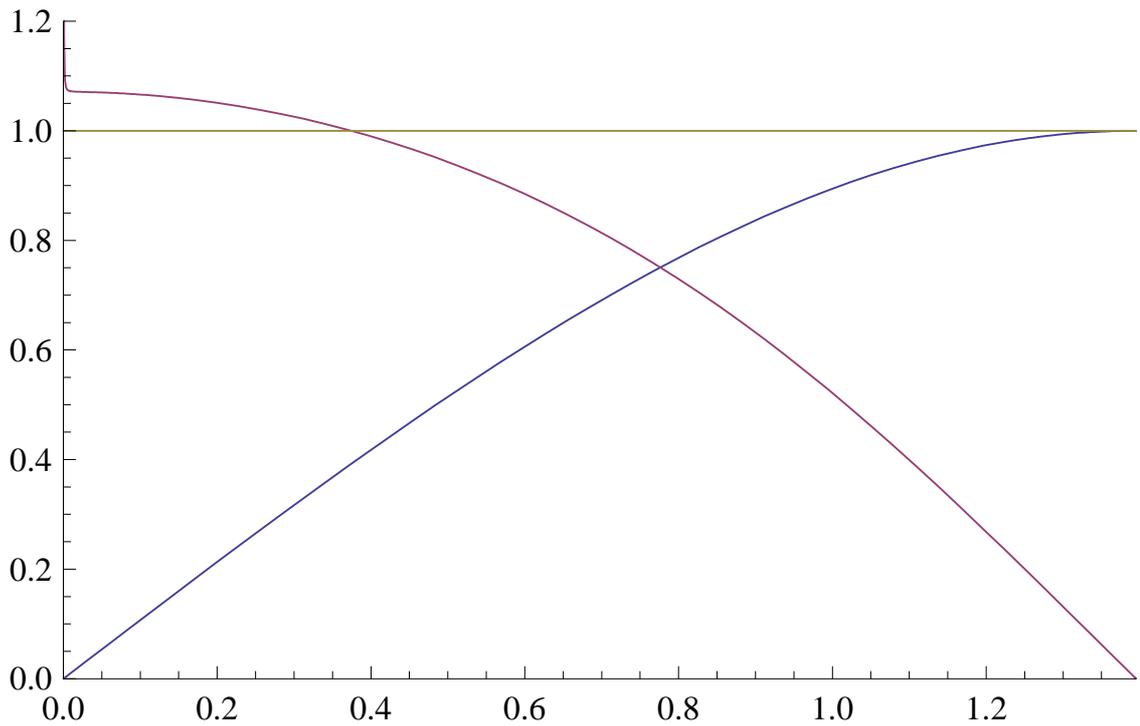

Figure 11: Non-standard gauge field kinetic term, and shooting from the boundary: for $e = 2.0$, and for the corresponding values $\lambda = 11.4515$ and $R = 1.394879$, the functions $f(r)$ and $f'(r)$ are shown. It is clearly seen that $f$, which starts at $f(R) = 1$, goes to zero at $r = 0$, whereas $f'$, which starts at $f'(R) = 0$, goes to some nonzero value which is not determined by the asymptotic analysis.

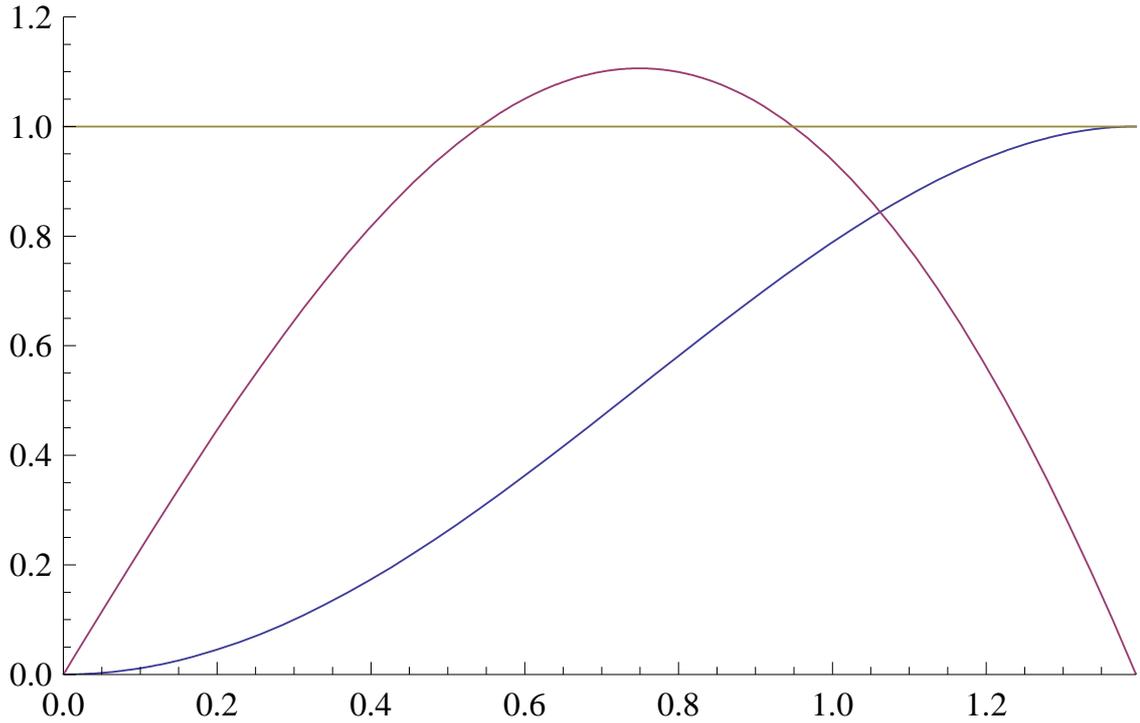

Figure 12: Non-standard gauge field kinetic term, and shooting from the boundary: for $e = 2.0$, and for the corresponding values $\lambda = 11.4515$ and $R = 1.394879$, the functions $\alpha(r)$ and $\alpha'(r)$ are shown. It is clearly seen that $\alpha$, which starts at $\alpha(R) = 1$, goes to zero at $r = 0$, and $\alpha'$, which starts at $\alpha'(R) = 0$, goes to zero, as well. Observe that the latter condition does not count as an independent boundary condition, because it is dictated by the symmetries of the corresponding differential equation (i.e., $\alpha$ has a power series expansion about $r = 0$ in terms of $r^2$ rather than $r$).